\documentclass[prd,twocolumn,showpacs,superscriptaddress,nofootinbib,floatfix]{revtex4-2}

\usepackage[utf8]{inputenc}
\usepackage{amsmath,amssymb}
\usepackage{slashed}
\usepackage{subfigure}
\usepackage[colorlinks=true, 
linkcolor=blue,
breaklinks=true,
urlcolor=magenta,
citecolor=blue]{hyperref}
\usepackage[usenames,dvipsnames]{color}
\usepackage{appendix}
\usepackage{braket,bm}
\usepackage{multirow}
\usepackage{enumitem}
\usepackage{color}
\usepackage{cancel}
\usepackage{orcidlink}
\usepackage{mathrsfs}

\usepackage{graphicx}
\usepackage{tikz}
\usepackage[normalem]{ulem}
\usepackage{booktabs}
\usepackage{array}
\usepackage{overpic}
\usepackage{float}
\usepackage{threeparttable}
\allowdisplaybreaks[4]

\graphicspath{{fig/}}

\usetikzlibrary{calc}
\usetikzlibrary{intersections}
\usetikzlibrary{trees}
\usetikzlibrary{decorations.pathmorphing}
\usetikzlibrary{decorations.markings}
\usetikzlibrary{arrows.meta}
\usetikzlibrary{patterns}
\tikzset{
   global scale/.style={
      scale=#1,
      every node/.append style={scale=#1}},
   photon/.style={decorate, decoration={snake}, draw=red},
   nucleon/.style={draw=black, postaction={decorate},
      decoration={markings,mark=at position .65 with{\arrow[draw=black]{latex}}}},
   pion/.style={draw=blue, postaction={decorate},
      decoration={markings,mark=at position .55 with{\arrow[draw=blue]{}}}},
    nucleonstar/.style={draw=black, postaction={decorate},
      decoration={markings, mark=at position 0.7 with {\arrow[draw=black]{latex}}}},
    }

\newcommand{\itp}{\affiliation{Institute of Theoretical Physics, Chinese Academy of Sciences, Beijing 100190, China}}

\newcommand{\su}{\affiliation{
Institute for Particle and Nuclear Physics, College of Physics,\\ Sichuan University, Chengdu, Sichuan 610065, China}}

\newcommand{\ucas}{\affiliation{School of Physical Sciences, University of Chinese Academy of Sciences, Beijing 100049, China}}

\newcommand{\phcc}{\affiliation{Peng Huanwu Collaborative Center for Research and Education, Beihang University, Beijing 100191, China}}

\newcommand{\hebtu}{\affiliation{Department of Physics and Hebei Key Laboratory of Photophysics Research and Application,\\
Hebei Normal University, Shijiazhuang 050024, China}}

\newcommand{\scnt}{\affiliation{Southern Center for Nuclear-Science Theory (SCNT), Institute of Modern Physics,\\
Chinese Academy of Sciences, Huizhou 516000, China}}

\newcommand{\md}{\mathrm{d}}

\begin{document}
\title{Rigorous Roy-Steiner equation analysis of $\pi K$ scattering at unphysical quark masses}

\author{Xiong-Hui Cao\orcidlink{0000-0003-1365-7178}}
\email{xhcao@itp.ac.cn}
\itp

\author{Feng-Kun Guo\orcidlink{0000-0002-2919-2064}}\email{fkguo@itp.ac.cn}
\itp\ucas\phcc\scnt 

\author{Zhi-Hui Guo\orcidlink{0000-0003-0409-1504}}\email{zhguo@hebtu.edu.cn}
\hebtu

\author{Qu-Zhi Li\orcidlink{0009-0001-2640-1174}}
\email{liquzhi@scu.edu.cn}
\su

\begin{abstract}
We perform a rigorous analysis of the $\pi K$ scattering at an unphysical pion mass 391~MeV using the Roy-Steiner equations, which satisfy unitarity, analyticity and crossing symmetry, for the first time. Stable solutions of the Roy-Steiner equations with different quantum numbers of isospin and angular momentum are obtained in the elastic energy region, by taking inputs from the $\pi K$ lattice data in the inelastic region, the lattice data from the crossed $\pi\pi \to K\bar{K}$ channels, the masses of $f_0(500)$ and $K^*$ at the same pion mass from previous study, and the Regge model. Predictions on the elastic $\pi K$ scattering phase shifts and the $K_0^*(700)$ pole content are made. Contrary to the virtual pole scenario obtained using the $K$-matrix method in the literature, we find that lightest strange scalar meson $K_0^*(700)$ remains a broad resonance at $m_\pi=391$~MeV. The cross-channel dynamics is found to play a crucial role in deriving the proper pole position. 
\end{abstract}

\maketitle

{\em Introduction.---}Understanding the spectrum of hadrons has been a fundamental challenge for decades because the underlying theory, quantum chromodynamics (QCD), is nonperturbative at low energies. 
In recent years, tremendous progress has been made using numerical lattice QCD in studying unstable hadron resonances that couple to two lower-lying hadrons (see recent reviews~\cite{Briceno:2017max, Mai:2021lwb}), and the first lattice calculation on three-body resonance $\omega$ has become available~\cite{Yan:2024gwp}. 
While many current lattice QCD calculations are performed at unphysically large quark masses due to computational resource limitations, this apparent drawback can also be seen as an opportunity. Such calculations provide a unique resource to explore the properties of hadron resonances at different quark masses, offering invaluable insights into the complex nonperturbative strong interaction dynamics that cannot be directly accessed through experiments~\cite{Hanhart:2008mx, Cleven:2010aw, Du:2017zvv,  Niehus:2020gmf, Molina:2020qpw}. 
Therefore it is definitely beneficial to fully exploit the lattice QCD results obtained at unphysical quark masses. 
However, a significant challenge lies in extracting the resonance properties from the lattice QCD data in a model-independent manner. Such extractions have usually been performed using the $K$-matrix method, which, while practical and satisfying unitarity, does not fulfill the analyticity and crossing symmetry requirements.
Our work addresses this issue by relying solely on fundamental properties of the scattering amplitudes, such as crossing symmetry, analyticity and unitarity, while using the currently available lattice data as inputs.
 
Over the past two decades, there has been a resurgence of interest in using dispersion relations to achieve high-precision determinations of observables in low-energy hadron physics.
Roy~\cite{Roy:1971tc} (for equal-mass scatterings) and Roy-Steiner (RS)~\cite{Hite:1973pm} (for unequal-mass scatterings) equations have been successfully applied to obtain precise scattering phase shifts and determine poles of the amplitudes for several key processes in the physical regime, such as the $\pi \pi$~\cite{Ananthanarayan:2000ht, Caprini:2005zr,Garcia-Martin:2011iqs, Moussallam:2011zg, Caprini:2011ky, Garcia-Martin:2011nna, Pelaez:2015qba}, $\pi K$~\cite{Buettiker:2003pp, Descotes-Genon:2006sdr, Pelaez:2020uiw, Pelaez:2020gnd} and $\pi N$~\cite{Hoferichter:2015hva,Cao:2022zhn, Hoferichter:2023mgy} scatterings. 
However, rigorous studies applying such frameworks in the context of lattice calculations with unphysical quark masses remain scarce. Notable exceptions include the first applications to $\pi \pi$ scattering in Refs.~\cite{Cao:2023ntr, Rodas:2023nec}. It should be emphasized that, compared to analyses in the physical regime, nontrivial modifications are often necessary when dealing with unphysically large quark masses. In such cases, resonances above the threshold in the physical situation can evolve into bound states below the threshold. 
For instance, the broad lightest scalar meson resonance $f_0(500)$, also known as the $\sigma$ meson, would become a bound state when the quark mass is large such that the pion mass $m_\pi=391~\mathrm{MeV}$~\cite{Briceno:2016mjc, Briceno:2017qmb}. This  necessitates the inclusion of a bound state pole term in the Roy equation formalism, which, however, is absent in the physical situation~\cite{Cao:2023ntr}.

In this Letter, we present the first RS equation analysis for the unequal-mass $\pi K$ scattering in light of recent lattice calculations. 
The task is technically much more challenging than the equal-mass $\pi \pi$ scattering due to the presence of complicated left-hand cuts and cross-channel dynamics. 
More importantly, a model-independent conclusion about the $K_0^*(700)$, also known as $\kappa$, resonance at unphysically large pion masses has not yet been reached when directly utilizing the currently available lattice results~\cite{ Wilson:2019wfr}. 
We will provide more constrained $\pi K$ scattering phase shifts and extract, in a model-independent way, the $\kappa$ pole content at $m_\pi=391$~MeV using the rigorous RS equation formalism. 
Our analysis reveals that the $K_0^*(700)$ remains a broad resonance at $m_\pi=391~\mathrm{MeV}$ when carefully incorporating lattice data from the $\pi\pi,K\bar K$~\cite{Dudek:2012gj, Dudek:2012xn, Briceno:2016mjc, Briceno:2017qmb} and crossed $\pi K,\eta K$~\cite{Dudek:2014qha,Wilson:2014cna, Wilson:2019wfr} channels by the Hadron Spectrum Collaboration (HSC) as inputs. 

{\em RS equation formalism.---}The $\pi K$ partial-wave (PW) amplitudes $f^I_J(s)$ with definite isospin $I$ and angular momentum $J$ satisfy RS equations, which are a set of crossing-symmetric PW dispersion relations, for which the Mandelstam variables $s$ and $u$ are constrained along the $su=b$ hyperbolae with $b$ a constant.
We have 
\begin{align}
   f_J^{I}(s)= &\, \frac{1}{\pi} \int_{m_{+}^2}^{s_\mathrm{m}} \mathrm{d} s^{\prime} \left\{K_{J, 0}^{I,\frac{1}{2}}\left(s, s^{\prime}\right) \operatorname{Im} f_{0}^{\frac{1}{2}}\left(s^{\prime}\right)\right.\nonumber\\
   & \left.+K_{J, 1}^{I,\frac{1}{2}}\left(s, s^{\prime}\right) \operatorname{Im} f_{1}^{\frac{1}{2}}\left(s^{\prime}\right)+K_{J,0}^{I,\frac{3}{2}}\left(s, s^{\prime}\right) \operatorname{Im} f_{0}^{\frac{3}{2}}\left(s^{\prime}\right)\right\} \nonumber\\
   & +\text{ST}^{I}_J(s)+\text{PT}^{I}_J(s)+s\text{DT}^I_J(s)+t\text{DT}^I_J(s) ,
   \label{eq:RS_s}
\end{align}
where $m_+^2\equiv (m_\pi+m_K)^2$, and $s_\text{m}$ denotes the matching point, above which external inputs are required. Crossing symmetry implies that the subtraction terms $\text{ST}^I_J(s)$  can be expressed in terms of the two $S$-wave scattering lengths $a^{1/2}_{0}$ and $a^{3/2}_{0}$. For unphysical pion masses around $400~$MeV, both the scalar meson $f_0(500)$ and the vector strange meson $K^*(892)$ become bound states below the $\pi\pi$ and $\pi K$ thresholds, respectively~\cite{Briceno:2016mjc,Wilson:2019wfr,Cao:2023ntr}, which should be explicitly included in the pole terms $\text{PT}^I_J(s)$~\cite{Cao:2023ntr} that would be absent in the physical situation. 
Particularly, crossing-symmetry dynamics requires to simultaneously include the direct $s$-channel $K^*(892)$, crossed $t$-channel $f_0(500)$, and $u$-channel $K^*(892)$ exchanges. Most previous works in the study of unphysical pion masses overlook this crossing symmetry requirement, which will be shown to be crucial for an accurate description. The $s$-channel driving terms $s\text{DT}^I_J(s)$ collect all the dispersion integrals over the high-energy tails, as well as higher PW contributions. 

Crossing-symmetric PW dispersion relations also need driving terms $t\text{DT}^I_J(s)$ from the crossed $\pi \pi \rightarrow K \bar{K}$ $t$-channel. The partial waves $g^I_J(t)$ for this coupled-channel system are typically modeled using the $K$-matrix approach, which satisfies unitarity but does not guarantee analyticity and crossing symmetry. 
The PW amplitudes $g^I_J(t)$ in the pseudo-physical region, which lies between the $\pi \pi$ and $K \bar{K}$ thresholds, are needed for the dispersive calculations. However, these amplitudes are quite challenging to obtain in lattice calculations and have not been available so far. 
Fortunately, the amplitudes $g^I_J(t)$ also fulfill the $t$-channel RS equations,
\begin{align}\label{eq:RS-t}
   g_{J}^I(t)=\Delta_{J}^I(t)+\frac{t}{\pi} \int_{4 m_\pi^2}^{t_\mathrm{m}} \frac{\md t^{\prime}}{t^{\prime}} \frac{\operatorname{Im} g_{J}^I(t)}{t^{\prime}-t},
\end{align}
where the integral describes the contributions from the $S$- and $P$-waves below the $t$-channel matching point $t_\text{m}$, above which high energy inputs are needed, and $\Delta_{J}^I(t)$ are the inhomogeneous terms including the subtraction constants, pole terms, the high-energy part of the $t$-channel, and the crossed $s$-channel inputs. 
Watson's final state theorem~\cite{Watson:1952ji} implies that the phase of $g^I_J(t)$ must coincide with the $\pi \pi$ phase shift in the pseudo-physical elastic region, which allows for a full reconstruction of the amplitudes using the Mushkelishvili-Omn\`es method~\cite{Muskhelishvili:1953, Omnes:1958hv}. 

Different from previous analyses for the physical situation~\cite{Buettiker:2003pp,Pelaez:2020gnd}, the presence of two bound state poles, $f_0(500)$ and $K^*(892)$, at the unphysically large pion mass $m_\pi=391$~MeV used in the lattice calculations necessitates modifications to the RS equations for both $\pi \pi \rightarrow K \bar{K}$ and $\pi K \rightarrow \pi K$. 
The Roy equation analysis of $\pi\pi$ scattering~\cite{Cao:2023ntr} at $m_{\pi} = 391~$MeV has provided a precise determination of the pole position and residue for $f_0(500)$. Additionally, the $K^*(892)$, a shallow vector bound-state at this pion mass, has been extensively studied in lattice QCD~\cite{Wilson:2019wfr}. 
We take the pole position as determined from the lattice study~\cite{Wilson:2019wfr}, while treating its coupling as an output of the RS equation analysis.

We need to incorporate particular driving terms that encompass the high-energy and high PW contributions above the matching point $s_\text{m}$, serving as inputs for the RS equations. These driving terms stem from lattice QCD calculations in Refs.~\cite{Dudek:2012gj, Dudek:2012xn, Briceno:2016mjc, Briceno:2017qmb}, as well as from the recent Roy equation analysis of $\pi\pi$ scattering at the same pion mass~\cite{Cao:2023ntr} and the Regge model~\cite{Veneziano:1968yb, Lovelace:1968kjy, Shapiro:1969km}.
The matching point is chosen to be $\sqrt{s_\text{m}}=1.30~\mathrm{GeV}$, above the $\eta K$ threshold $1.14~$GeV but below the three-body $\pi\pi K$ threshold $1.33~$GeV. 
In order to obtain a rigorous and stable solution, we follow the procedure in Refs.~\cite{Hoferichter:2015hva,Buettiker:2003pp, Cao:2023ntr} by defining a $\chi^2$-like object between the input and the output of each dispersion relation at many different energy points below $s_\text{m}$, which is then minimized numerically. 
Technical details can be found in an accompanying longer article~\cite{Cao:2025hqm}.

\begin{figure}[tbh]
   \centering
   \includegraphics[width=.45\textwidth,angle=-0]{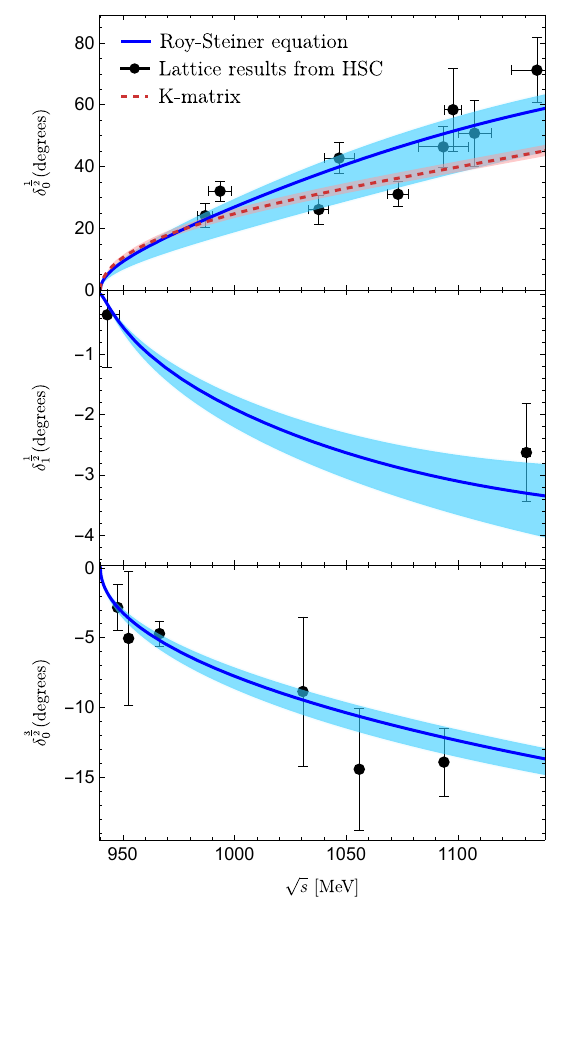} 
   \caption{The $(I,J)=(\frac{1}{2},0), (\frac{1}{2},1), (\frac{3}{2},0)$ phase shifts at $m_\pi=391$~MeV obtained by solving the RS equations. The solid lines and shaded bands represent the central values and  uncertainties. The lattice QCD results from HSC~\cite{Wilson:2014cna} are also shown for comparison. We also show as a red dashed curve the $(\frac{1}{2},0)$ phase shift from the preferred $K$-matrix parametrization in Ref.~\cite{Wilson:2014cna}.} \label{fig:s_PW}
\end{figure}
{\em Results.---}After successfully obtaining stable solutions $f^I_J(s)$ of the $\pi K$ RS equations, the phase shifts $\delta^I_J(s)$ can be obtained using $f^I_J(s)=\sin\delta^I_J(s)e^{i\delta^I_J(s)}/\rho(s)$, valid below the inelastic $\eta K$ threshold, where $\rho(s)\equiv \sqrt{\left(s-m_+^2\right)\left(s-(m_\pi-m_K)^2\right)/s}$ is the phase space factor. 
The most relevant $S$- and $P$-wave $\pi K$ phase shifts from RS equations are plotted in Fig.~\ref{fig:s_PW}, together with the lattice results from Ref.~\cite{Wilson:2014cna} obtained with a rather distinct way via the L\" uscher formula~\cite{Luscher:1990ux}. 
The excellent agreement with the lattice data in the elastic region is a nontrivial test of our solution of the RS equations.
Given that the most relevant channel for extracting the $\kappa$ pole comes from the $S_0^{1 / 2}$ wave, we also compare the phase shift from the $\pi K$-$\eta K$ coupled-channel $K$-matrix parametrization fit preferred in Ref.~\cite{Wilson:2014cna}, which incorporates a single pole coupled to both channels plus a constant background matrix. 
While our dispersive solution tends to be slightly larger than the $K$-matrix result above 1~GeV, the HSC results fall within the uncertainties of our solution.

We emphasize that the phase shifts below the matching point $s_\text{m}$ derived from the RS equations are pure predictions, rigorously considering crossing symmetry and accounting for various uncertainties. 
The uncertainties include those from 
the lattice QCD data of the $\pi K$ scattering at and above $s_{\mathrm{m}}$~\cite{Wilson:2014cna,Wilson:2019wfr}, and the crossed $t$-channel $S$- and $P$-wave phase shifts of $\pi\pi$ scattering~\cite{Cao:2023ntr},
the high energy and the higher PW dynamics parameterized by Regge models, 
the $K^*(892)$ pole position $\sqrt{s_{K^*}}=(934\pm 2)~\text{MeV}$~\cite{Wilson:2019wfr}, and the $\sigma$ pole position, $\sqrt{s_\sigma}=759^{+~7}_{-16}~\text{MeV}$, and its coupling to pions, $g_{\sigma\pi\pi}=493^{+27}_{-46}~\text{MeV}$~\cite{Cao:2023ntr}. 
The bootstrap method is employed to propagate uncertainties of the lattice and Regge inputs to the solutions of the RS equations. 
A series of pseudo-inputs are generated by randomly varying the lattice data within their errors, and the RS equations are solved for each pseudo-input.

Among these sources of uncertainty, the primary ones stem from those of the lattice QCD data for $\pi K$ scattering at and above $s_\text{m}$, which account for nearly all errors except those in the $P$-wave. The $P$-wave is strongly influenced by the pole position of the near-threshold bound state $K^*(892)$. In contrast, the uncertainties from the $f_0(500)$ pole and the $t$-channel $S$- and $P$-wave phase shifts have only a mild impact on the $s$-channel $\pi K$ scattering amplitude. 
The model dependence of the asymptotic Regge amplitude has an almost negligible influence on the low-energy phase shifts. 

\begin{figure}[tb]
   \centering
   \includegraphics[width=.45\textwidth,angle=-0]{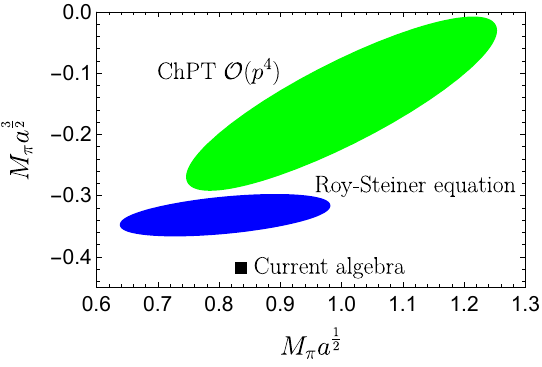} 
   \caption{Comparison between the determination from solving the RS equations and the NLO ChPT calculations, as well as the current algebra result at $m_\pi=391$~MeV. The green ellips correspond to the error estimations for the ChPT $\mathcal{O}(p^4)$ results.} \label{fig:SL_plot}
\end{figure}
The scattering lengths defined as $a_0^I=\frac{2}{m_{+}} f_0^I\left(m_{+}^2\right)$ from RS equations at $m_\pi=391$~MeV are obtained as
\begin{align}
   m_\pi a_0^{\frac{1}{2}}=0.92^{+0.06}_{-0.28},\quad m_\pi a_0^{\frac{3}{2}}=-\left(0.32^{+0.05}_{-0.02}\right), 
\end{align}
between which there is a moderate correlation, as represented in Fig.~\ref{fig:SL_plot}. 
Predictions of $\pi K$ scattering lengths from chiral perturbation theory (ChPT) at the next-to-leading order (NLO)~\cite{Bernard:1990kw, GomezNicola:2001as} with the low-energy constants (LECs) from Ref.~\cite{Bijnens:2014lea}, are also plotted for comparison. Note that we have neglected the correlations among the LECs, which might cause an overestimate of the uncertainties. 
Our main focus here, however, is on the size of the ellipse region and its possible overlap with the results of the RS equations. 
Interestingly, the ellipse representing the NLO chiral results does not overlap with those derived from the RS equations. 
We also show the leading order ChPT (current algebra) result. The notable discrepancy between the RS result and the ChPT ones indicates that the convergence of SU(3) ChPT is questionable at $m_\pi = 391~$MeV.

The $P$-wave phase shift with $(I,J)=(1/2,1)$ in Fig.~\ref{fig:s_PW} turns out to be negative, which clearly indicates the bound-state nature of the $K^*(892)$ at $m_\pi=391$~MeV, where we have used the convention that phase shift equal to zero at the threshold.  
Meanwhile, an associated virtual state pole is also found at nearly the same location on the second Riemann sheet of the complex $s$ plane.
This result is consistent with previous HSC analyses~\cite{Dudek:2014qha, Wilson:2014cna, Wilson:2019wfr}. 

The $I=\frac{3}{2}$ $S$-wave phase shift of is similar to the experimental one in the physical situation, both of which are dominated by non-resonant repulsive interactions.

\begin{figure}[tb]           
   \centering 
   \includegraphics[width=.5\textwidth,angle=-0]{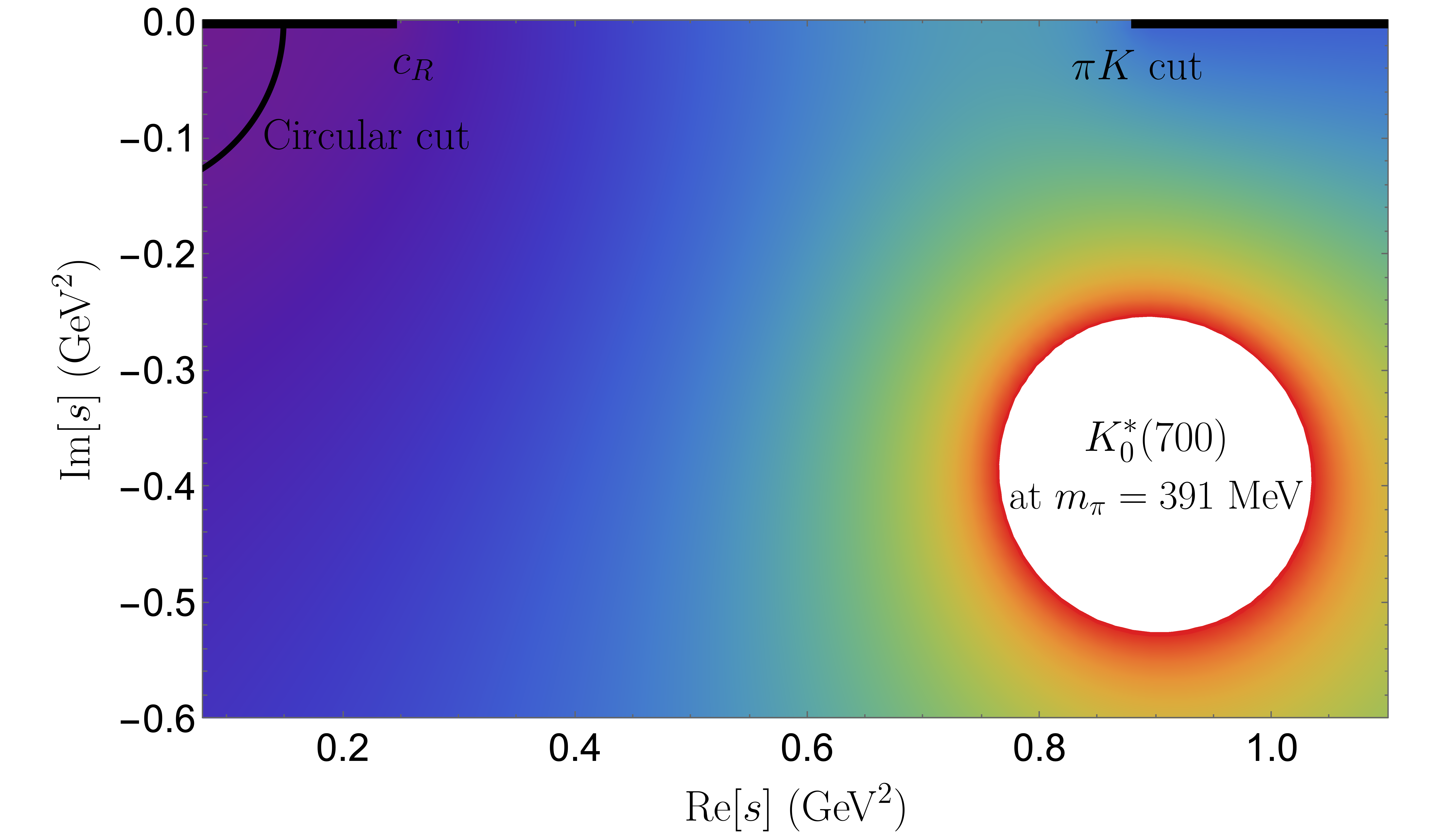} 
   \caption{Moduli of the $S$-matrix $S^\frac{1}{2}_0(s)=1+2i\rho(s)f^\frac{1}{2}_0(s)$ on the second Riemann sheet computed from the RS equations at $m_\pi=391$~MeV, where the $K_0^*(700)$ resonance pole is clearly revealed.  The unitarity $\pi K$ cut along the real axis, the left-hand cuts and a circular cut are depicted as black lines.} 
   \label{fig:kappa} 
\end{figure}

The $I={1}/{2}$ $S$-wave phase shifts, depicted in Fig.~\ref{fig:s_PW}, rise slowly and do not cross 90 degrees in the elastic region, indicating the absence of a standard narrow-width resonance in this channel, at least in the elastic region. 
The pole content of the amplitude in this case is subtle. It was shown in Refs.~\cite{Zheng:2003rw, Yao:2020bxx} that deep virtual state poles and broad resonance poles can yield similar slow-growing positive phase shifts. Indeed, a deep virtual state pole below the $\pi K$ threshold was reported in the $K$-matrix analysis in Ref.~\cite{Dudek:2014qha}. 
However, the $I={1}/{2}$ $\pi K$ $S$-wave in all our solutions of the RS equations has a very broad $\kappa/K_0^*(700)$ pole, as shown in Fig.~\ref{fig:kappa}. 
The pole position and residue are
\begin{align}
\sqrt{s_\kappa} & =966_{-24}^{+41}-i 198_{-17}^{+38}~{\rm MeV}\,, \nonumber \\
g_{\kappa} & =\left(759^{+63}_{-25}\right)e^{-i \left(1.05^{+0.14}_{-0.06}\right)}~{\rm MeV}\,.
\end{align}
One notices that, when the unitarized ChPT (UChPT) amplitudes~\cite{GomezNicola:2001as} have their parameter freedom constrained by the finite-volume spectra presented in Refs.~\cite{Dudek:2014qha, Wilson:2014cna}, a pole with a real part close to the threshold and a large imaginary part was also obtained in Ref.~\cite{Wilson:2019wfr}. 
Since the next-to-leading order UChPT amplitudes have certain left-hand cut contributions from the $t$- and $u$-channel one-loop diagrams for the interaction kernel, the qualitative similarity in the pole position obtained using such amplitudes with ours can be regarded as one more evidence for the importance of the left-hand cuts in the pole extraction.
However, the UChPT amplitudes are known to exhibit violations of crossing symmetry since the resummation is performed only in the $s$-channel.

The pole position can be compared with the pole obtained in Ref.~\cite{Pelaez:2020gnd} at $(678\pm 7)-i (280\pm 16)$~MeV in the physical situation.
The comparison indicates that the $K_0^*(700)$ remains a pole with a large imaginary part for pion masses ranging from the physical value up to at least 391~MeV. 
A fast transition, in which the resonance $K_0^*(700)$ evolves into a pair of virtual state poles, is expected to occur at a pion mass larger than 391~MeV, in contrast to its SU(3) counterpart, $f_0(500)$.\footnote{Detailed Roy equation analyses demonstrate that the $f_0(500)$ evolves into a pair of subthreshold virtual state poles at $m_\pi\simeq300$~MeV~\cite{Rodas:2023nec}.} Here the $t$-channel $f_0(500)$ pole plays a crucial role.

The drastic difference between the $K_0^*(700)$ pole positions at $m_\pi=391$~MeV obtained using the $K$-matrix method in Ref.~\cite{Dudek:2014qha} and the RS equations in this work can be attributed to the different treatments of the cross-channel dynamics. 
In the $K$-matrix analysis, the influence of the left-hand cut was neglected~\cite{Dudek:2014qha, Wilson:2014cna, Wilson:2019wfr}. 
While this approximation may be reasonable in the physical situation, it becomes inadequate for unphysically large quark masses, especially when the particles in the crossed channels become bound states. 
Indeed, at $m_\pi=391$~MeV, both the $f_0(500)$ and $K^*(892)$ become bound states, and new left-hand cuts below the $\pi K$ threshold, from the $t$-channel $f_0(500)$ and $u$-channel $K^*(892)$ exchanges, appear in the $S$-wave $\pi K$ amplitude. 
The nearest branching point to the threshold is $c_R=\frac{1}{2}\left(2m_\pi^2+2m_K^2+\sqrt{\left(4m_\pi^2-s_\sigma\right)\left(4m_K^2-s_\sigma\right)}\right)$, arising from the $t$-channel $f_0(500)$ exchange. 
Taking $\sqrt{s_\sigma}=759~$MeV~\cite{Cao:2023ntr}, the distance between the branching point $c_R=491~$MeV to the virtual state pole location of $600 \sim 750~$MeV reported in Ref.~\cite{Dudek:2014qha}, is similar to that between the pole and the threshold $m_+=940~$MeV. 
In this case, the left-hand cut from the $t$-channel $f_0(500)$ exchange is crucial for a precise determination of the $K_0^*(700)$ pole position. 

As a further check, by excluding the pole terms in the crossed channels and all other non-Cauchy off-diagonal integral contributions from crossing symmetry in the $I={1}/{2}$ $S$-wave RS equation~\eqref{eq:RS_s}, we find two virtual state poles below the threshold---one far from the threshold and the other near it: $\sqrt{s_\text{vs}}=312$ and $910~$MeV. The solution from this incomplete treatment is qualitatively similar to the findings using the $K$-matrix approach~\cite{Dudek:2014qha} that also neglects the left-hand cuts. 
Therefore, our findings highlight the importance of considering cross-channel dynamics to achieve a stable and accurate determination of pole positions, especially when resonances are located far from the physical energy region. 
Consequently, this serves as a critical reminder to be cautious about the convergence range of SU(3) ChPT and the key role of the cross-channel dynamics, such as when using unitarized ChPT~\cite{Nebreda:2010wv} to investigate the light quark mass dependence of the $K_0^*(700)$ and other similar broad resonances.

{\em Summary.---}In this Letter, we have presented results from the RS equations for the $\pi K$ scattering at unphysically large pion mass of 391~MeV. 
The extension of the RS equations from the physical situation to such a large pion mass is nontrivial, as the bound state poles of the $\sigma$ and $K^*(892)$ mesons must be included in the pole terms.
Crossing symmetry implies that these bound state poles also result in additional relevant cuts of the partial wave scattering amplitudes. 
We have provided the first model-independent results for $\pi K$ scattering at $m_{\pi}=391$~MeV, utilizing the currently available lattice data beyond the elastic energy region and Regge theory estimates for higher partial waves and in the higher energy region. 
The obtained $\pi K$ scattering phase shifts in the elastic region agree well with the results from the unconstrained $K$-matrix parameterization fit~\cite{Dudek:2014qha, Wilson:2014cna} within the uncertainties. Nevertheless, the $K$-matrix method does not have reliable left-hand cuts, which are crucial to precise determination of the $K_0^*(700)$ pole position in the continuum.

We found that the $K_0^*(700)$ state at $m_\pi=391~$MeV does not manifest as a deep virtual state pole below the $\pi K$ threshold as reported in the $K$-matrix method~\cite{Dudek:2014qha}, but rather remains as a broad resonance similar to the experimental one. We have found that a deep virtual state pole emerges when all left-hand cuts from crossed channels are ignored. Obviously, this indicates that the cross-channel dynamics cannot be overlooked when analyzing the broad $\kappa/K^*_0(700)$ resonance. 

The rigorous RS equation analysis can be generalized to other systems at unphysical quark masses, such as the scattering between pions and heavy mesons or $\pi N$ scattering, once lattice QCD results become available.
New insights into the resonance structures and the connection between these resonances and chiral symmetry can be expected from such studies.

\bigskip 

\begin{acknowledgements}
We thank Zhi-Guang Xiao and Han-Qing Zheng for useful discussions.
This work was supported in part by National Natural Science Foundation of China (NSFC) under Grants No.~12347120, No.~12475078, No.~12150013, No.~12125507, No.~12361141819, and No.~12447101 and No.~12335002; by the Postdoctoral Fellowship Program of China Postdoctoral Science Foundation under Grant No.~GZC20232773 and No.~2023M74360; by Chinese Academy of Sciences under Grant No.~YSBR-101; and by the National Key R\&D Program of China under Grant No.~2023YFA1606703. 
ZHG is also partially supported by the Science Foundation of Hebei Normal University with Contract No.~L2023B09. 
\end{acknowledgements}

\bibliography{refs}
\end{document}